# Automated Optical Reading of Scanned ECGs

Manuel Pazos-Santomé[1], Fernando Martín-Rodríguez[2](*), Mónica Fernández-Barciela[2].

[1]*Biomedical Engineer, University of Vigo.*

[2]*atlanTTic research center for Telecommunication Technologies, University of Vigo, C/ Maxwell SN (Campus LagoasMarcosende), 36310 Vigo, Spain*

E-mail: fmartin@uvigo.es

**Abstract**: Electrocardiogram (ECG) is a valuable tool for medical diagnosis used worldwide. Its use has contributed significantly to the prevention of cardiovascular diseases including infarctions. Although physicians need to see the printed curves for a diagnosis, nowadays there exist automated tools based on machine learning that can help diagnosis of arrhythmias and other pathologies, these tools operate on digitalized ECG data that are merely one-dimensional discrete signals (a kind of information that is much similar to digitized audio). Thus, it is interesting to have both the graphical information and the digitized data. This is possible with modern, digital equipment. Nevertheless, there still exist many analog electrocardiogram machines that plot results on paper with a printed gris measured in millimeters. This paper presents a novel image analysis method that is capable of reading a printed ECG and converting it into a sampled digital signal.

## 1. Introduction

According to the World Health Organization (WHO), ischemic heart disease is the leading cause of death worldwide and has experienced a drastic increase in the last 20 years [1]. Nowadays, there are tests that contribute to the prevention of the disease, among them, the electrocardiogram (ECG). ECG information can be plotted on a paper (analog ECG) or digitalized to produce sampled discrete signals very similar in form to digital audio.

The aim of this work is to allow reading ECG's available only in paper format, converting them into digital ones. Paper ECGs must be optically scanned and processed by a novel image analysis application.

In the available literature, there are few examples of a system created for this same purpose. As examples, we found the systems described in [2] and [3].

In [2], authors rely on grayscale thresholding to remove the millimeter grid from the scanned images. This technique seems less robust that the colorimetry-based method presented in this work. Application in [2] also integrates an OCR for reading of patient data, that can be useful avoiding manual typing that is always prone to human errors.

In [3], authors use again grayscale techniques to perform grid extraction. Besides they incorporate interesting features like correcting rotated images and they assess their system correlating digitalized data with directly sampled signals. This means that authors have used images from "modern" digital equipment that produce both the images and the digital signals. Although results are good, tests should be made also with analog plotted ECG's.

Main interest of this kind of systems is the post processing of digitalized ECG data in automated applications for helping diagnosis. This kind of applications usually use machine learning. Many examples can be found in the literature. For example, in [4], Liu et al, publish a review on the application of deep learning for ECG diagnosis. As another example, in [5], authors use a classical multi-layer perceptron (MLP) to detect arrhythmic EGC's.

## 2. Materials and Methods

### 2.1. Scanned ECG's

Plotted ECGs are scanned using a standard flatbed scanner. For appropriate working, scanner must be configured in color mode and 600 dpi resolution.

Obtained images look like the example in Figure 1.

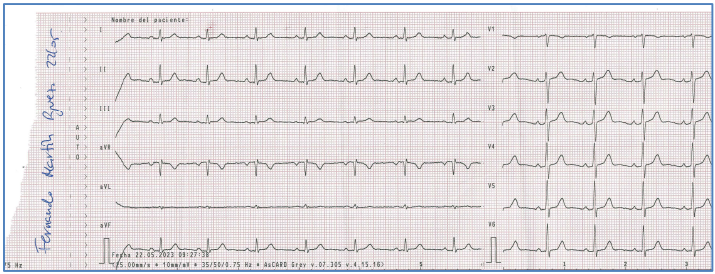

Fig. 1. Example of scanned ECG.

Sample ECGs are donated by volunteers and when necessary, they have been manually anonymized by, simply, over impressing a white rectangle to hide the name. This task has been done using image edition software.

### 2.2. Application flow diagram

A flow diagram depicting the processing stages of the application is drawn in figure 2. Blue blocks are preprocessing ones, while marron ones are character removing blocks (a special necessary stage), light green blocks are about curve tracking, finally, black green blocks are devoted to numeric signals creation.

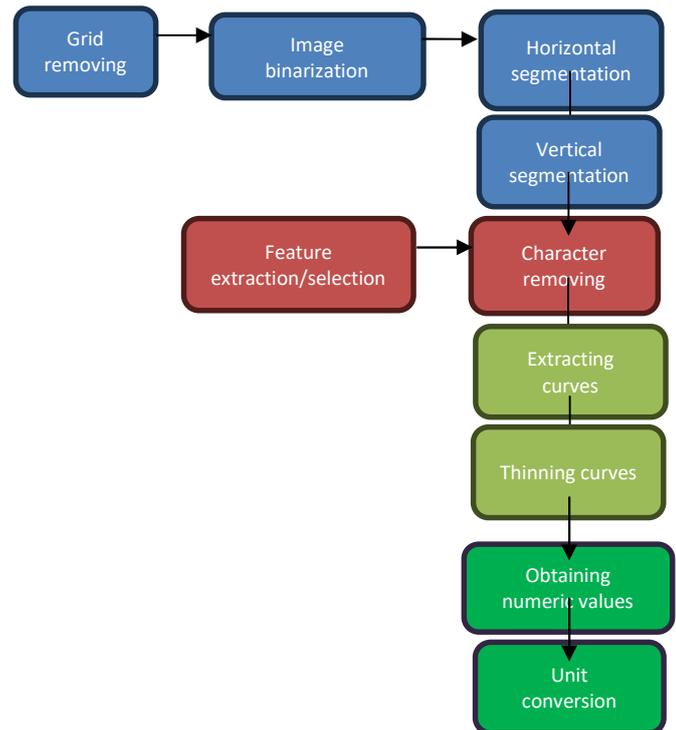

Fig. 2. Application flow diagram.

In the next subsections, the different stages are described.

### 2.3. Grid removing and binarization

The first stage in Figure 2 is removing the millimeter grid. As seen in Figure 1, human eye distinguishes it properly due to its different color. Analyzing at low level, this "pink like" color can be easily characterized as having a stronger red component in the RGB space, figure 3.

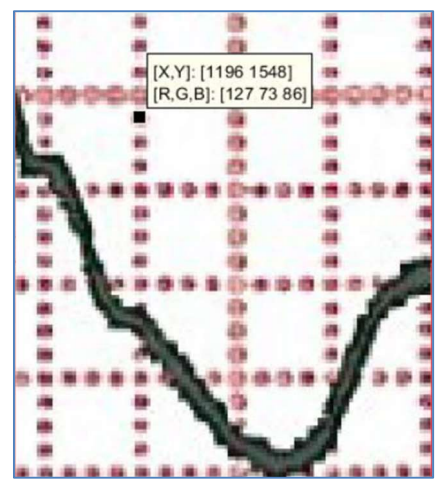

Fig. 3. Application flow diagram.

In these conditions, a normalized color index can be used. Color indexes are extensively used in remote sensing application over hyperspectral images. For example, NDVI index (Normalized

Differential Vegetation Index) can detect healthy plants (rich in chlorophyll) by computing a simple equation **NDVI=(Nir-Red)/(Nir+Red)** [6]. Big reflectance of Nir (Near infrared component) due to chlorophyll is the basis for this index.

Generally speaking, a normalized index is defined as **idx=(C₁-C₂)/(C₁+C₂)**, where **C₁** and **C₂** are two color components. If **C₁** dominates over **C₂** for the interest regions, **idx** will yield high values (note that **idx** is always in the -1.0, +1.0 range).

In this case, input image is RGB with only three color channels. For detecting the grid, the next index has been defined:

$$idx = \frac{Red - \frac{1}{2}Green - \frac{1}{2}Blue}{Red + \frac{1}{2}Green + \frac{1}{2}Blue} \qquad (1)$$

Where **C₂** has been substituted by the average between the two weal channels (Green and Blue).

After computing the index, this new image (the index image) is binarized to obtain a binary mask. To obtain this mask, the binarization procedure is as follows:

1. Index image has negative pixels, these ones are "rectified" using a ReLU function: ReLU(x)=(x+|x|)/2.
2. Otsu threshold is computed (Otsu threshold [7] is widely used in OCR technology and computer vision). For this problem, a scaling (50% of Otsu threshold) is necessary.

Mask obtained with this method is shown in Figure 4.a. This mask is applied by adding it to the grayscaled version of the input image (Figure 4.b).

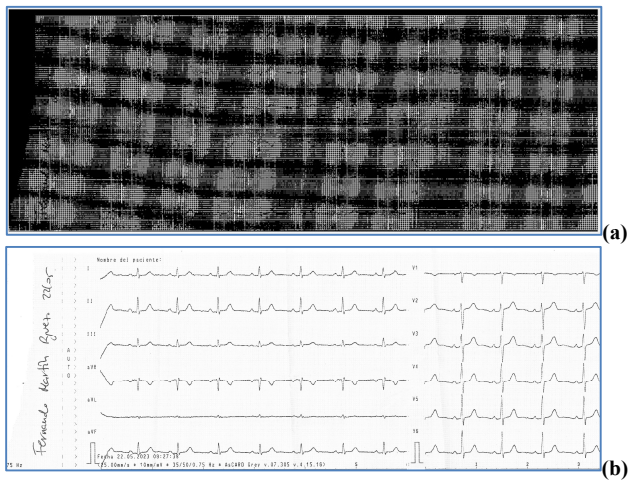

Fig. 4. (a) binarized mask, (b) addition of mask and grayscaled input image.

Image in figure 4.b (grayscale input without gird) is binarized using again Otsu threshold. In this case, it is computed for the new image and used without modification. After binarization, work image is complemented so that active color is white (or true as it is a logical image), Figure 5.

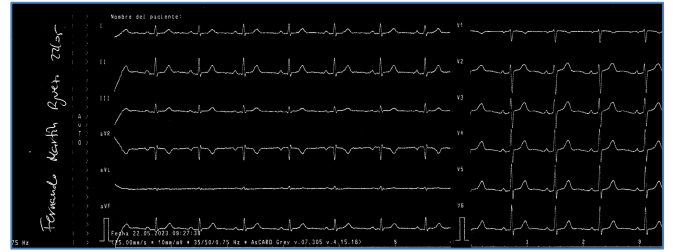

Fig. 5. Input image binarized with grid removed.

## 2.4. Horizontal and vertical segmentation

Segmentation can be defined as the extraction of the interest parts of an image for further processing. In this case, horizontal segmentation consists of determining the initial and ending lines of each ECG curve. Output of horizontal segmentation would be a collection of sub-images consisting of stripes of the image in figure 5.

Vertical segmentation is performed on the stripe images. The purpose now is finding the initial and ending column of each curve. Defining the left curves as the most important ones, output of this second segmentation would be a new collection of smaller stripes containing only the curves.

Horizontal segmentation is based on vertical projection, i.e. computing the sum of each line, a local minimum should be obtained between each two curves. In order to make projection values independent of image size, mean line value is used instead of sum. See a real projection as a signal waveform in Figure 6. Local maxima are first detected as points where value is greater than its neighborhood. The minima between each two maxima define the "breaking lines" for segmentation. A graphical depiction of process can be found in Figure 6. Find one of the strip images in Figure 7.

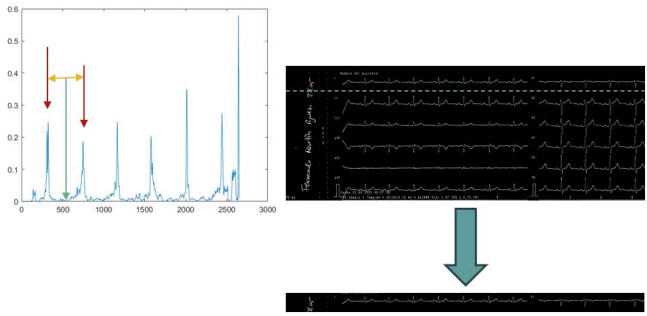

Fig. 6. Vertical projection.

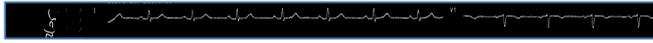

Fig. 7. Input image binarized with grid removed.

Vertical projection is run on the stripe images (like Figure 7). In this case, another kind of projection is used. Instead of summing or averaging pixel values, the number of edges is counted column wise (edge: change from black to white or vice versa). Segmentation points (columns) can be detected, searching small interval where this projection vanishes. The last interval in the left part and the first one in the middle region define the region of interest, see Figure 8.

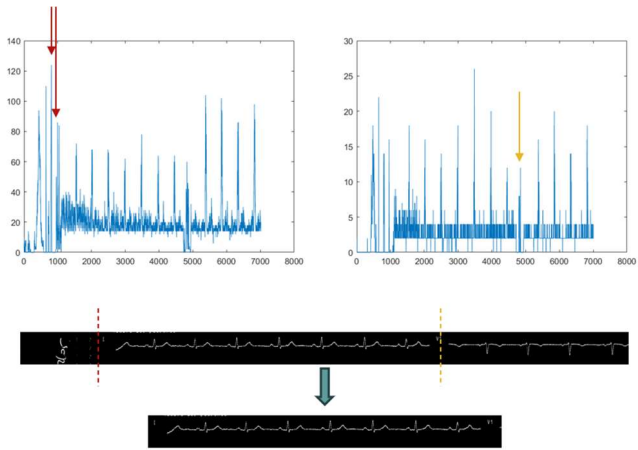

Fig. 8. Vertical segmentation.

## 2.5. Removing characters

This stage is necessary due to the printed characters present in the lowest curve (aVF graph). These characters are still present in the segmented version of this curve, Figure 9.

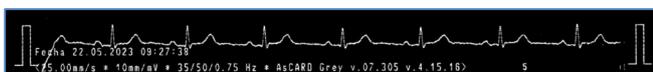

Fig. 9. Segmentation of aVF curve.

At this moment, characters are small white blobs that, very likely, could be detected by simple

geometric properties: width, height, aspect ratio, y-coordinate of the centroid…

To avoid selecting small strokes of the curves, that could be separated from the main line, it is recommendable to apply a preprocessing consisting of a morphological close [8]. Recommendable structuring element is a small circle. This operation will join together the broken pieces. Circle radius cannot be very high because it would cause characters to get connected to the curve.

After closing the image, a labeling operation can be done to detect all connected blobs and computing their geometric properties. To select the most appropriate numeric features, Pearson correlation coefficient [9] was used, measuring significance of each parameter. From this analysis, aspect ratio and character y-coordinate yield the most significant ones. Filtering blobs by comparing these values with the average of a real characters, the curve gets cleaned, see Figure 10.

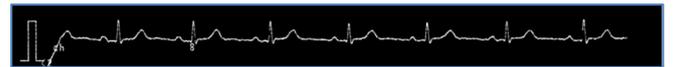

Fig. 10. aVF curve after removing characters.

## 2.6. Extracting and thinning curves

Working images at this point should be a set of curves, already binarized and segmented but not necessarily continuous. Very frequently they are broken (vanishing) mostly in the signal maxima. Many times, images also present small noise blobs and/or artifacts. For a proper tracking and digitalization, curves should be connected, cleaned and assured to have 1 pixel thickness (thinned).

To accomplish the cleaning task, a process based on mathematical morphology is designed. Process is as follows:

1. First, a close operation removes the breaks. Now the structuring element is big to assure curves pieces to get joined. This operation also causes that the maxima and the minima regions be converted into solid objects. See Figure 11.

2. Image is labeled and only the biggest blob is maintained.

3. This image is used as a mask and multiplied by the original one (Figures 8 and 10) resulting in a cleaned image.

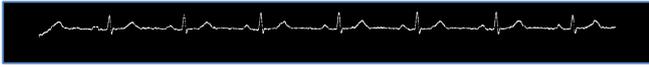

Fig. 11. Cleaned curve.

The curve in Figure 11 is re-segmented vertically so that curve is tightly bounded by image left and right limits. Note that curve breaks are still there.

Thinning is done now in a simple manner, creating a new curve, where the active points are taken from the first active point found at each column (tracking each column from top to bottom). Experimentally, this method worked better than other more elaborated methods; for example, morphological thinning produced a lot of artifacts.

As it will be seen in the next subsection, these last images are already good to obtain numerical signals.

## 2.7. Obtaining numerical signals

To obtain a numerical signal, process is simply the creation of two vectors. The first one will contain, initially, the horizontal coordinates of active points. Afterwards, this vector will be converted into a time base. Initially, this vector contains integer consecutive numbers from 0 to the image width minus one.

The second vector must contain a vertical coordinate per column (vector size must be equal to that of the time base). Using the last image from last subsection, we should have only one active point per column, of course the y-coordinate of that point is the numeric value to be saved.

A problem arises when there is a break in the curve. Signal value would not be defined for some columns. In this case, the missing values are computed as the median value for the valid points in an interval centered on each missing point.

Having both vectors full of values, note that these will be coordinate values, units will be pixels for both. Desired units would be seconds (or milliseconds) for the time base and volts (or mV) for the amplitude values.

This operation should be easy knowing the values of milliseconds per millimeter and millivolts per millimeter. These values are printed in the ECG sheet but, in our case, are introduced manually by user (for our examples 0.1 mV/mm and 0.04 ms/mm).

Of course, it is needed to know the spatial resolution of the image (dpi: dots per inch or millimeters per pixel: mm/px). Although scanning resolution is known, it was decided to obtain this value from the image, i.e. the number of pixels between two consecutive lines/columns of the grid can measured. This is easily done using horizontal and vertical projections of the grid mask (Figure 4.a). This process is illustrated in figure 12.

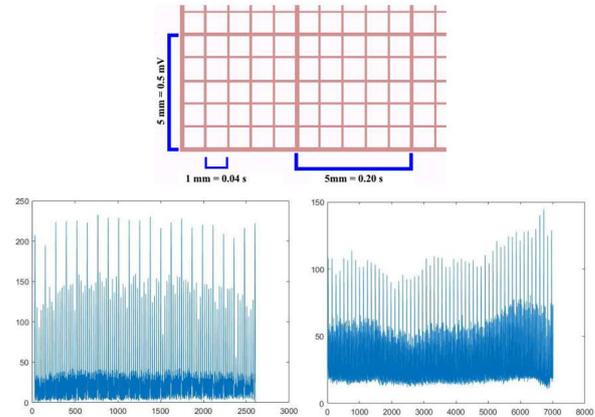

Fig. 12. Computation of spatial resolution.

Once vectors are in the proper units, amplitude vector is "centered" by subtracting its own median value. This is done because original curves have no reference for 0 Volt value. Time bases will start at zero with no problem, but for the amplitude values a median of zero seems a reasonable election (assuming zero value when signals are at rest state).

## 3. Results and discussion

### 3.1. Graphical application

Code for all methods has been implemented in MATLAB [10]. A graphical application has been developed to ease user experience, Figure 12. This MATLAB app can be exported as a `.exe` file that can be executed outside of MATLAB (installing a public runtime module).

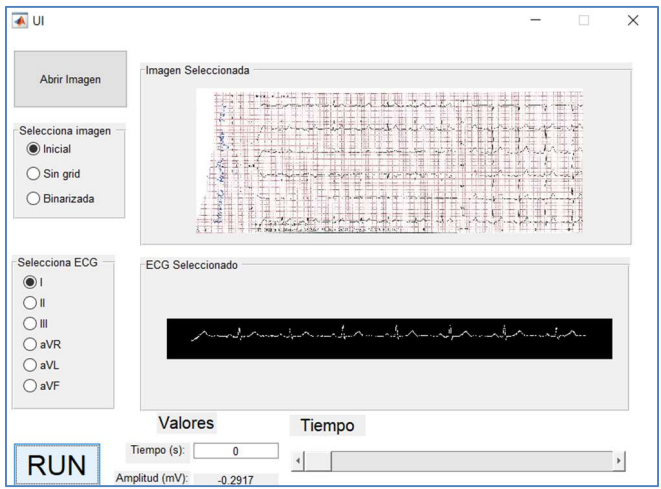

Fig. 13. Computation of spatial resolution.

## 3.2. Stress test

Results were appropriate for all the images tested. A stress test was performed, analyzing the performance against distorted images. Results were:

- Worst distortion is rotation. Even a small rotation angle makes system to fail.
- System resists low intensity blur.
- System resists low intensity desaturation.
- System resists the combined action of low intensity blur and desaturation.

## 4. Conclusions

A useful system for digitalization of plotted electrocardiograms has been developed and tested. Results are good as long as images are of acceptable quality. Some future lines of work can be point at:

- Obtaining a more appropriate dataset of scanned ECG's to test system more thoroughly. Particularly, testing with plots with other formats (this probably will make necessary to create other preprocessing versions).
- Correct image distortions (tilt, blur, and desaturation) before processing.
- Improve character removing. Even recognizing them to obtain important information (specially values for unit conversion that nowadays are introduced manually).
- Testing other methods for curve tracking taking vanishing points.

## Acknowledgments

Authors have received no funding for this research. Authors acknowledge their institutions: University of Vigo and AtlantTIC for their support and for the necessary resources supplied.

## Conflict of interests/Competing interests

The authors declare that they have no conflict of interests and no competing interests.

## References

[1] https://www.who.int/es/news-room/fact-sheets/detail/the-top-10-causes-of-death, last seen on 12th of August, 2024.

[2] L. Ravichandran, C. Harless, A. J. Shah, C. A. Wick, J. H. Mcclellan and S. Tridandapani, "Novel Tool for Complete Digitization of Paper Electrocardiography Data," in IEEE Journal of Translational Engineering in Health and Medicine, vol. 1, pp. 1800107 -1800107 , 2013, Art no. 1800107, doi: 10.1109/JTEHM.2013.2262024.

[3] Julian D. Fortune, Natalie E. Coppa, Kazi T. Haq, Hetal Patel, Larisa G. Tereshchenko, "Digitizing ECG image: A new method and open-source software code," Computer Methods and Programs in Biomedicine, Volume 221 (2022), Art no. 106890, doi: 10.1016/j.cmpb.2022.106890.

[4] Xinwen Liu, Huan Wang, Zongjin Li, Lang Qin, "Deep learning in ECG diagnosis: A review," Knowledge-Based Systems, Volume 227 (2021), Art no. 107187, doi: 10.1016/j.knosys.2021.107187.

[5] A. Vishwa, M. Lal, S. Dixit, P. Vardwa, "Clasification Of Arrhythmic ECG Data Using Machine Learning Techniques," International Journal of Interactive Multimedia and Artificial Intelligence, Volume 1(4), pp 67-70 (2011), doi: 10.9781/ijimai.2011.1411.

[6] https://www.usgs.gov/landsat-missions/landsat-normalized-difference-vegetation-index, last seen on 12th of August, 2024.

[7] N. Otsu, "A Threshold Selection Method from Gray-Level Histograms," in IEEE Transactions on Systems, Man, and Cybernetics, vol. 9, no. 1, pp. 62-66, Jan. 1979, doi: 10.1109/TSMC.1979.4310076.

[8] González, R.C.; Woods, R.E.; Eddins, S.L. Digital Image Processing Using MATLAB; Gatesmark Publishing, Cop.: Upper Saddle River, NJ, USA, 2009.

[9] Tadist, K., Najah, S., Nikolov, N.S. et al. "Feature selection methods and genomic big data: a systematic review." J Big Data, Volume 6 (2019), Art no. 79, doi: 10.1186/s40537-019-0241-0.

[10] https://es.mathworks.com/products/matlab.html, last seen on 12th of August, 2024.